\documentstyle[12pt,aasms4]{article}
\input psfig.sty

\def\lea{\mathrel{<\kern-1.0em\lower0.9ex\hbox{$\sim$}}}
\def\gea{\mathrel{>\kern-1.0em\lower0.9ex\hbox{$\sim$}}}

\lefthead{The authors}
\righthead{The RGB tip at the center of M31}

\begin{document}

\title{No Increase of the Red-Giant-Branch Tip \\ Luminosity Toward the
       Center of M31}
\author{P. Jablonka}
\affil{DAEC-URA173, Observatoire de Paris-Meudon, Place Jules Janssen,
F-92195 Meudon, France }

\author{T.J. Bridges\altaffilmark{1}}
\affil{Institute of Astronomy, Madingley Road, Cambridge, UK, CB3 0HA}

\author{A. Sarajedini\altaffilmark{2}}
\affil{Department of Physics and Astronomy, San Francisco State
University, 1600 Holloway Ave, San Francisco, CA 94132, USA}

\author{G. Meylan}
\affil{European Southern Observatory, Karl-Schwarzschild-Strasse 2,
D-85748 Garching-bei-M\"unchen, Germany}
 
\author{A. Maeder, G. Meynet}
\affil{Geneva Observatory, CH-1290 Sauverny, Switzerland}

\altaffiltext{1}{previously Royal Greenwich Observatory, Madingley
Road, Cambridge CB3 0EZ, UK}
\altaffiltext{2}{Hubble Fellow}


\renewcommand{\thefootnote}{\fnsymbol{footnote}}

\begin{abstract}

We present observations with the Hubble Space Telescope Wide Field
Planetary Camera 2 of three fields centered on super-metal-rich
globular clusters in the bulge of M31. Our $(I,V-I)$ color-magnitude
diagrams reach as faint as $I$ $\sim$ 26.5 mag and clearly reveal the
magnitude of the first ascent red giant branch (RGB) tip. We find that
the apparent $I$ magnitude of the RGB tip does not become brighter
near the center of M31 as concluded by previous investigators.  Our
observations and artificial star experiments presented in this
study strongly support the idea that previous very bright stars were
likely the result of spurious detections of blended stars due to
crowding in lower resolution images.  On the contrary, our
observations indicate that, at a mean projected galactocentric
distance of 1.1 kpc, the RGB tip is some 1.3 magnitudes {\it fainter}
than it is at 7 kpc. An analysis of this difference in RGB tip
magnitude suggests that the M31 bulge stellar population has a mean
metallicity close to that of the Sun.

\end{abstract}


\section{Introduction}

For more  than a decade, the  nature of the stellar population towards
the center of M31 has been the subject of considerable debate.  In the
first such studies using modern CCD detectors,  Mould (1986) and Mould
\& Kristian (1986)  observed   four fields located on   the South-East
minor  axis  of  M31,  at 20~kpc,  12~kpc,  7~kpc,  and 5~kpc from the
center,  respectively.   Rich  \&  Mighell (1995) subsequently reached
more   central regions,  into  the    inner  500~pc  of the    galaxy.
Summarizing their  own   work based  on  WFPC1  data  along with  some
previous work  based on ground-based  data,   they confirmed that  the
first-ascent red   giant  branch (hereafter  referred  to  as RGB) tip
seemed to brighten towards the center  of M31.  At  7 kpc, the RGB tip
appears to  have the   same luminosity  as that observed   for typical
Galactic globular  clusters  ($I$ $\sim$  20.5~mag  at the distance of
M31)  while it becomes significantly brighter  near the center of M31.
Rich  \& Mighell (1995)  mentioned the  apparent contradiction between
such  a RGB tip luminosity brightening   and its predicted theoretical
dimming, due to a  decrease in luminosity with increasing metallicity,
as expected  when  going from the   halo to the more  metal-rich bulge
stellar population.  See, e.g., Bica et al.  (1990) for the metal-rich
nature of the M31 bulge.  As also illustrated  by Bica et al.  (1991),
higher metallicity populations normally  have fainter RGBs, due to TiO
blanketing   in  the  $I$-band.    In order   to   account   for their
observations, Rich \& Mighell (1995)   proposed that the M31 bulge  is
younger than the Galactic halo by 5 to 7~Gyr and/or  the presence of a
rare stellar evolutionary phase.

Renzini (1993, 1998) discussed in  detail the simpler possibility that
the apparent brightest RGB stars  in the bulge  of M31 were merely the
result of image  crowding, and concluded that  this was indeed likely.
Further support for this conclusion came from the work of Depoy et al.
(1993) who obtained ground-based $K$-band,  $2.2 \mu$m photometry of a
604  arcmin$^2$  field in   Baade's Window   and  carried out  careful
artificial star experiments.  They simulated the appearance of Baade's
Window at the distance of M31  and obtained an artificial brightening,
due to  crowding, of more  than  1 mag,  reproducing the most luminous
stars found by Rich \& Mould (1991) and Davies et al. (1991).

This letter  presents the  analysis of   our HST  WFPC2  observations.
These high-resolution images in  the central part  of the bulge of M31
provide us   with   an  ideal   means  by  which   to  construct  deep
color-magnitude diagrams (CMDs) of  the  stellar populations near  the
center of this  galaxy.  In addition,  they allow us to quantitatively
assess and take into account the degree of crowding.

\section{Observations and Data Reduction}

We obtained Hubble Space Telescope WFPC2 images during  Cycles 5 and 6
with the F555W  ($V$) and F814W ($I$) filters;  our targets were three
fields centered on super-metal-rich star clusters  in the bulge of M31
(Jablonka et al. 1992).  Two fields, around the star clusters G170 and
G177, are located SW along the major  axis of M31, respectively at 6.1
and 3.2 arcmin  from the galaxy  nucleus; the third field, around  the
cluster G198, is located NE along the major axis at 3.7 arcmin (Huchra
et al.  1991).  Adopting 1 arcmin = 250~pc, as Rich \& Mighell (1995),
these  angular separations correspond to  projected distances of about
1.55,  0.80, and 0.92  kpc, respectively.   In  Figure~1, we show  the
location of the Rich \& Mighell (1995) fields (squares) and our fields
(circles).  The  region shown  is a   13.5   $\times$ 13.5  arcmin box
centered on  the M31 nucleus,  and the  squares/circles are  $\sim$ 80
arcsec in size/diameter.

Each cluster was centered on the PC  chip and exposed for 4800 seconds
in $I$ and 5200 seconds  in $V$, in a  series of 1100- and 1300-second
subexposures.

After   standard   recalibration   of  our    data,   we   used    the
DAOPHOT/ALLSTAR/ALLFRAME software package for crowded-field photometry
(Stetson  1994), along  with  the PSFs kindly  provided  to us  by the
Cepheid-Distance-Scale  HST key project.  The F555W and
F814W instrumental magnitudes were {\it  in fine} converted to Johnson
$V$ and Cousins $I$ magnitudes, using the zero  points and color terms
given by Hughes   et al.  (1998).    See Jablonka et  al.  (1999)  for
further details of our observations and data reduction.

We have not applied any aperture corrections  to our photometry. There
are simply no sufficiently bright,  isolated stars in our very crowded
frames to  allow us to obtain our  own aperture corrections. It is not
possible  to use aperture  corrections  from other WFPC2 data, because
there  are  variations of  the aperture   corrections  with  time (HST
breathing, etc; see  Suchkov  \& Casertano 1997).  However, previously
published  studies  based  on  the same   PSFs   as those used  herein
(e.g. Holland et  al.  1996; Holland et  al.  1997; Hill et  al. 1998;
Hughes  et  al. 1998)  find that the  aperture  corrections are rarely
larger than $\sim$0.05 mag. In addition,  our own photometric study of
the M31 globular cluster G1 (Jablonka et al. 1999; Meylan et al. 1999)
confirms this fact.  As a result, we estimate  that our  assumption of
zero aperture correction introduces  an uncertainty of $\pm$ 0.05  mag
into our photometry.

\section{Results and Discussion}

We  constructed a ($I$,$V-I$)   CMD for  each   of our three   fields.
Between $\sim$35,000 and    $\sim$55,000 stars were detected in    the
Planetary Camera,  depending on the field.  The  high density of stars
superposed  upon  the bright,   unresolved  bulge background  in these
fields    prevents the  detection    of  faint   stars. This   induces
incompleteness due to a rather bright detection limit.

Figure~2 shows the CMD of the field around  the cluster G170 in the PC
chip.   The mean  photometric errors,  representing the frame-to-frame
dispersion in the measured magnitudes, are indicated; these correspond
to  0.3 mag at  $I=26$ mag, 0.08  mag  at $I=24$  mag and  0.05 mag at
$I=22$ mag.  A  total of 53,036 stars are  detected in  this field.  A
red  clump at $I$ = 24.5--25.0  mag can be  seen in Figure~2, although
rather  hidden  by the  high density of   points; it  is  more clearly
visible in  the luminosity function.  Similar red  clumps, at the same
$I$  mag, are   observed in the  M31  globular   cluster G1 (Rich   et
al. 1996; Jablonka et al.   1999).  The CMDs of  our other two  fields
are  similar to the one presented  in Figure~2:  all have well defined
red giant branches.

Figure~3 displays the mean locii of the CMDs for our three bulge
fields (PC frames). The points were obtained by splitting the data in
intervals of 0.65 mag in $I$ and 0.4 mag in $V-I$. In each interval,
the mean value was calculated using the Numerical Recipes function
{\tt moment} which gives a proper measure of the absolute deviation of
a distribution. The resulting mean locii are clearly the signature of
very metal-rich stellar populations (cf. Fig. 1 of Bica et al. 1991),
as is the red clump seen in Figure~2.  Within the errors, the three
fields exhibit the same mean stellar population.  No gradient of the
stellar population is expected, given our photometric precision.

We estimate, from Figure~3, the apparent $I$ band magnitude of the RGB
tip of the  mean stellar population.  This yields  $I(tip) = 22.39 \pm
0.38$ mag  for the G170 field,  $I(tip) = 22.31  \pm 0.44$ mag for the
G177 field, and $I(tip) = 22.27 \pm 0.48$ mag for the G198 field.  The
uncertainties in  these  quantities are the  results  of combining, in
quadrature,  the error in the measurement  of this quantity along with
an estimated error  of $\pm 0.05$ mag  in the aperture corrections and
$\pm   0.05$ mag  in  the    Holtzman  et   al.  (1995)    photometric
transformations.  Schlegel et al.  (1998) find an average reddening of
E($B-V$) = 0.062 towards  M31.   Applying E($V-I$) = 1.3~E($B-V$)  and
A$_I$ = 1.46~E($V-I$),  and  adopting the Cepheid distance  modulus of
24.43 mag (Freedman \& Madore 1990), we obtain a mean location for the
RGB tips in our three fields of M$_I$  $\sim$ $-2.5 \pm 0.4$ mag; this
places the  mean  metallicity of  these fields  between  those of  the
Galactic bulge globular clusters NGC~6553/NGC~6528 and Terzan 1, if we
follow the ranking  of  Bica  et al.  (1991).  Interpreted  within the
context of the abundances published   by Barbuy et al.  (1998),   this
suggests that the  mean M31 bulge  metal abundance in  these fields is
approximately solar.

In order to allow for the comparison  of our results with the previous
work  summarized by Rich  \&  Mighell (1995),  we have determined  the
magnitude   of   the upper   {\it  envelopes} of    the   RGBs in  our
color-magnitude diagrams.\footnote{ According to an anonymous referee,
the  quantity measured by Rich  \& Mighell is  more closely akin to the
upper envelope of the RGB rather than the RGB tip.}  We find $I(env) =
21.0 \pm 0.2$ mag for the G170 field, $I(env) = 20.8  \pm 0.2$ mag for
the G177 field,  and $I(env) = 20.9  \pm 0.2$ mag  for the G198 field.
Although these values  seem to indicate  a slight brightening  towards
the galaxy  center, we do not  consider this significant, as {\it (i)}
the values are the same within the error bars, and {\it (ii)} crowding
becomes  severe with  decreasing  radius  and  induces  an  artificial
brightening as will be described below.

In Figure~4, the  crosses indicate the  previous determinations of the
apparent  RGB upper envelope as a  function of  the projected distance
(in  kpc) from  the  nucleus of  M31.  All of   these values come from
ground-based and pre-refurbishment   HST  WFPC1 data.  Note  that  the
point at 7  kpc is taken  from the work  of Mould \&  Kristian (1986).
The three filled hexagons,  presented with their  error bars,  are our
new results.   We see that our PC values are $\sim$1.5 mag fainter
than   WFPC1 and  ground-based   measurements  at  similar  radii.  We
interpret this  difference as due  to severe  crowding problems in all
previous  data at these radii.  A  similar conclusion  was reached  by
Grillmair et  al.  (1996)   in  their WFPC2  study  of  the M32  bulge
population.  They did not find the  very luminous AGB stars present in
previous ground-based studies, and experiments with their WFPC2 frames
degraded to ground-based spatial resolution showed  that many of these
`stars' are  in fact the  result of image blends  due to crowding.  We
are confident that  the  point  at 7  kpc has  not  been significantly
affected  by  crowding  as we note   that   Mould \&  Kristian  (1986)
performed  their  stellar photometry using   two techniques - aperture
photometry and PSF fitting. They point out that their results were not
significantly affected by the choice of reduction method. We take this
as suggestive evidence indicating that crowding was not a problem.

We  ran numerous  artificial star  experiments  in order to  check the
validity  of our photometry  in  all  of  our fields.   We limit  this
description to the tests performed on the PC and the WF2 frames of the
globular  cluster G170.  The WF2 frame  is selected among the three WF
frames, since, because of its orientation, the mean surface brightness
in this frame  is closest to  that observed in the  PC frame; WF2 thus
provides a good location to isolate and analyse the effect of a change
in spatial resolution (pixel  size). We added 210  stars to the PC and
419   stars to the  WF2  frames,  all of   them distributed  along the
fiducial RGB as shown in Panel (a) of  Figure~5. From those stars, 193
(91\%) are recovered in the PC frame and 310  (73\%) in the WF2 frame.
Panels (c)  and (d) of Figure~5  display the resulting  photometry. In
the PC frame, the input sequence is very well recovered, even though a
small (constant) error shallows  it a bit.  However,  this is  not the
case for the WF2 frame, where  a clear spurious increase in luminosity
is detected, an effect  which is larger  for fainter magnitudes as can
be seen in  panel (b) of Figure~5.  This  induces, along the y-axis, a
shrinking of the color-magnitude  diagram with a general shift towards
brighter magnitudes. In a forthcoming paper, we will discuss at length
the fact that this crowding effect in the WF2 frame, due to a degraded
spatial resolution when  going from the PC  frame to the WF2 frame, is
similar  to keeping the  PC resolution but observing increased stellar
density. In their artificial star experiments,  Rich \& Mighell (1995)
take as input a narrow stellar luminosity distribution, centered on an
already rather bright magnitude.    Thus, as the genuine bright  stars
are hardly affected by the crowding, they  indeed recover their stars.
However, their   experiments do not test  the  hypothesis that fainter
stars  (not  existing  in   their sample)  would create   some  bright
agglomerates in  their frames.  From  our own  experiments, it becomes
clear that only  the PC frames  can be reliably  used for the study of
stellar populations in the M31 bulge.

As described above, the RGB tip becomes fainter  as one approaches the
nucleus of M31, from a  value of I$\sim$20.5--21  mag at a radius of 7
kpc  to  I$\sim$22.3 mag  inside of 2  kpc.   The work of  Da Costa \&
Armandroff (1990) showed  that the first ascent RGB  tip  of a typical
Galactic globular with $[Fe/H]<-0.7$ has  $M_I$ $\sim$ $-4.0$ mag. For
an age larger than      about 8~Gyr and    abundances in   the   range
$-1.7<[M/H]<-0.7$, the  isochrones     of Bertelli   et   al.   (1994)
corroborate the  assertion of Da  Costa \& Armandroff (1990) regarding
the  constancy of  the  $I$-band  RGB tip luminosity  at  $M_I$ $\sim$
$-4.0$   mag.  Adopting the M31  distance  modulus and reddening given
above, this translates to I$\sim$20.5 mag, which is similar to the RGB
tip of M31 at 7 kpc. The simplest conclusion we can make based on this
apparent consistency is that the M31  stellar population at 7 kpc from
the   center is similar   in  metallicity  and  age  to the   Galactic
globulars.

Earlier, we showed  that the M31 bulge at  a mean distance of 1.1  kpc
from that galaxy's center has an RGB tip  magnitude that is consistent
with  an old, approximately solar  metal abundance stellar population.
Thus, our results confirm the steep metallicity  increase from halo to
bulge  stars in   M31  and, given  the  probable   uncertainty in  our
metallicity estimate, are  in-line  with the analysis of  the spectral
features of the M31 nucleus by Bica et al. (1990).

\section{Conclusions}

Since none of our observations probe the nucleus of M31, there remains
the possibility that a rare stellar population could be present in the
nucleus as claimed by Rich \& Mighell (1995; see also Davidge et al.
1997); however, in our view, the results of our observations and
analysis make this possibility unlikely.

The observations presented in this study strongly support the idea
that very bright stars were likely the result of spurious detections
of blended stars due to crowding in WFPC1 and ground-based images, as
also suggested by Renzini (1993, 1998), Depoy et al.  (1993), and
Grillmair et al. (1996).  Only the refurbished HST with the WFPC2
Camera can cope with projected stellar densities as high as 55,000
stars within a 36$^{\prime\prime}$ $\times$ 36$^{\prime\prime}$ area.
   
In addition, based upon the absolute $I$ magnitude of the first ascent
RGB tip, we  conclude that the  M31 bulge at  7 kpc from the center is
consistent  with  an   old  intermediate-to-metal-poor  (i.e. $[Fe/H]$
$\lea$ --0.7) population.   This is in  stark contrast with the  bulge
stars inside 2 kpc from the center, which are also old but have a mean
metal abundance that is approximately solar.

\acknowledgements

Ata Sarajedini   was supported by the  National  Aeronautics and Space
Administration   (NASA)  grants HF-01077.01-94A,  GO-05907.01-94A, and
GO-06477.02-95A from the Space  Telescope Science Institute, which  is
operated by the Association of Universities for Research in Astronomy,
Inc., under NASA contract NAS5-26555.

\begin{figure}[h]
\centerline{
\psfig{bbllx=10mm,bblly=48mm,bburx=200mm,bbury=250mm,%
figure=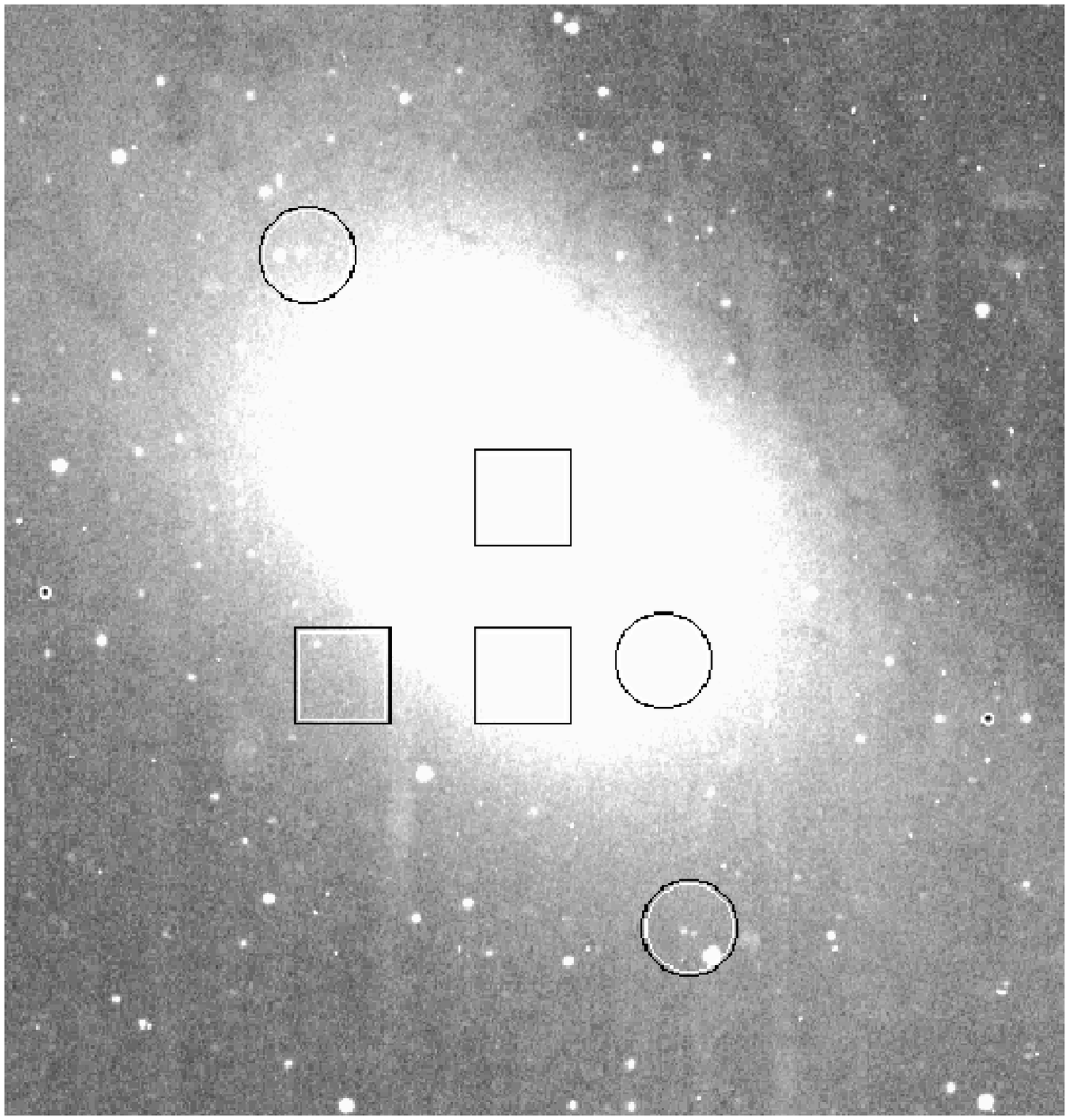,width=\the\hsize}}
\caption[]{\label{fig.diagram}  The  location of  the M31 bulge fields
observed with   the  Space Telescope.   Rich   \&  Mighell  (1995) are
represented with squares while the   fields analysed in this work  are
shown with circles.   The region shown is a  13.5 $\times$ 13.5 arcmin
box centered on the M31 nucleus, and the squares/circles are $\sim$ 80
arcsec in diameter/size.  North is to the top, East to the left.  }
\end{figure}

\begin{figure}[h]
\centerline{
\psfig{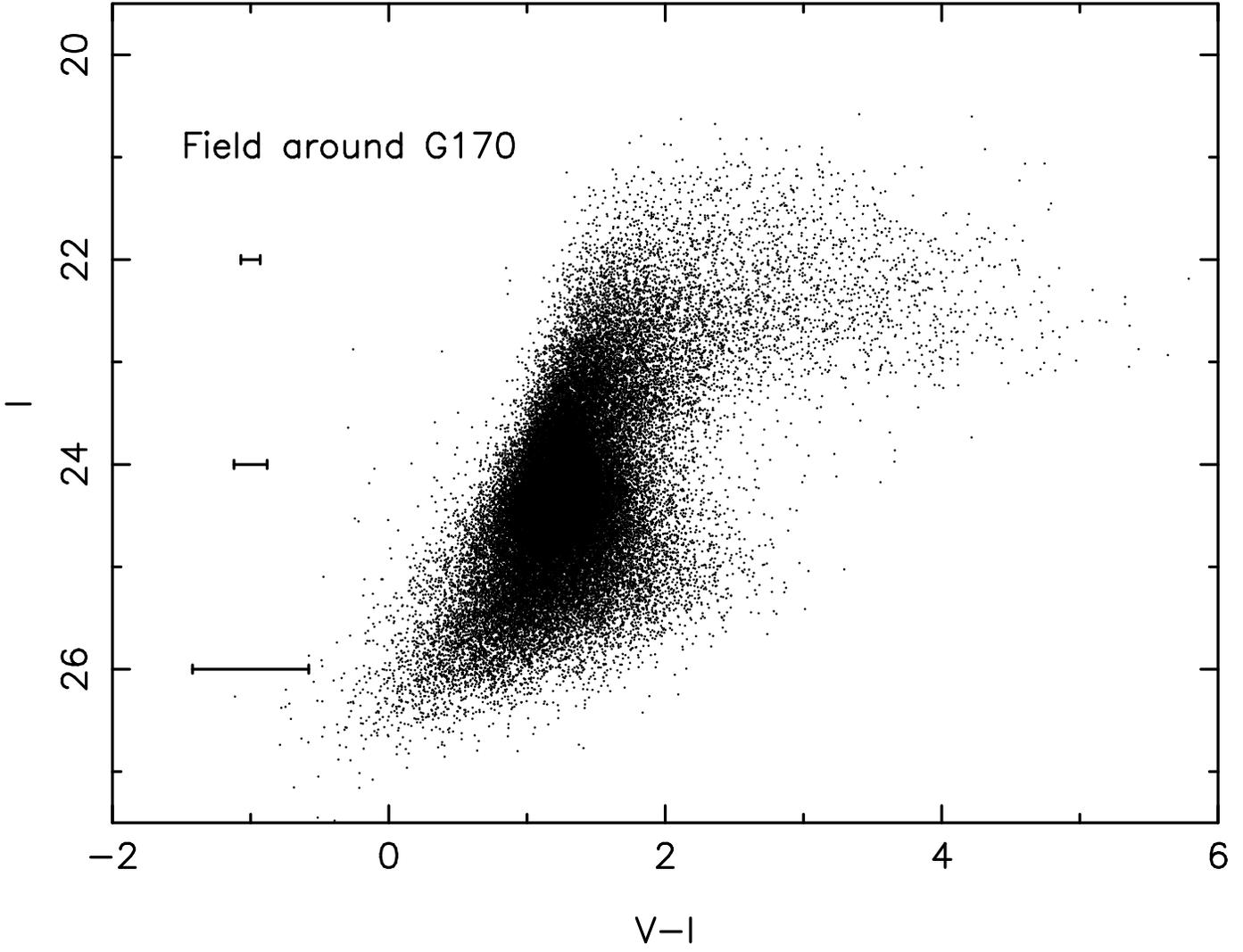}}
\caption[]{\label{fig.diagram} The Color-Magnitude  diagram of  the PC
field stars around the cluster G170}
\end{figure}

\begin{figure}[h]
\centerline{
\psfig{bbllx=23mm,bblly=70mm,bburx=182mm,bbury=228mm,%
figure=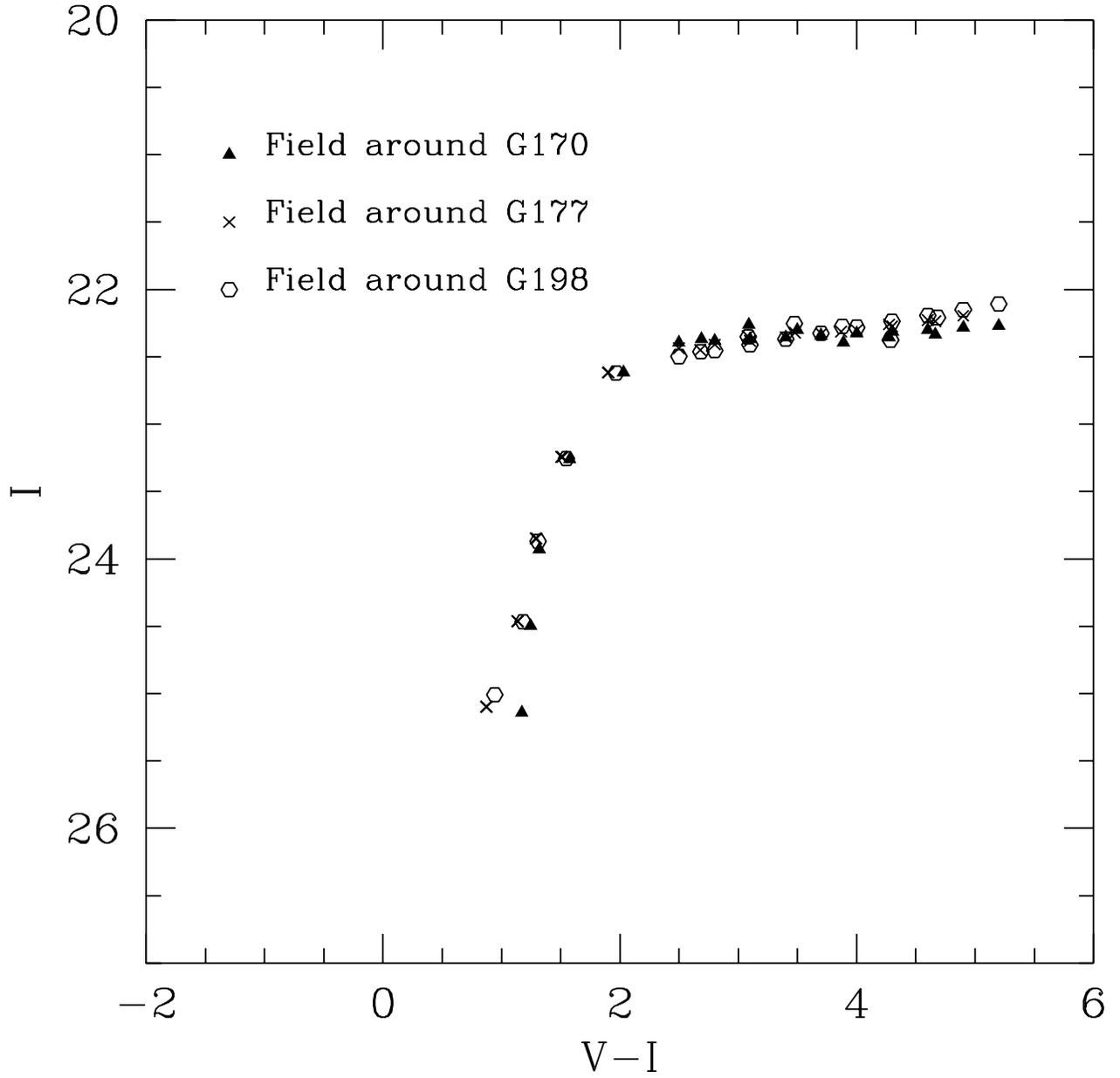,width=\the\hsize}}
\caption[]{\label{fig.diagram} The mean  locii of the  color-magnitude
diagrams  in the  PC frames  for  our  fields  around  the three  star
clusters. Each field is represented by a  distinctive symbol. The star
clusters themselves  are not considered  here, only the  field stellar
population.  }
\end{figure}

\begin{figure}[h]
\centerline{
\psfig{bbllx=41mm,bblly=85mm,bburx=174mm,bbury=228mm,%
figure=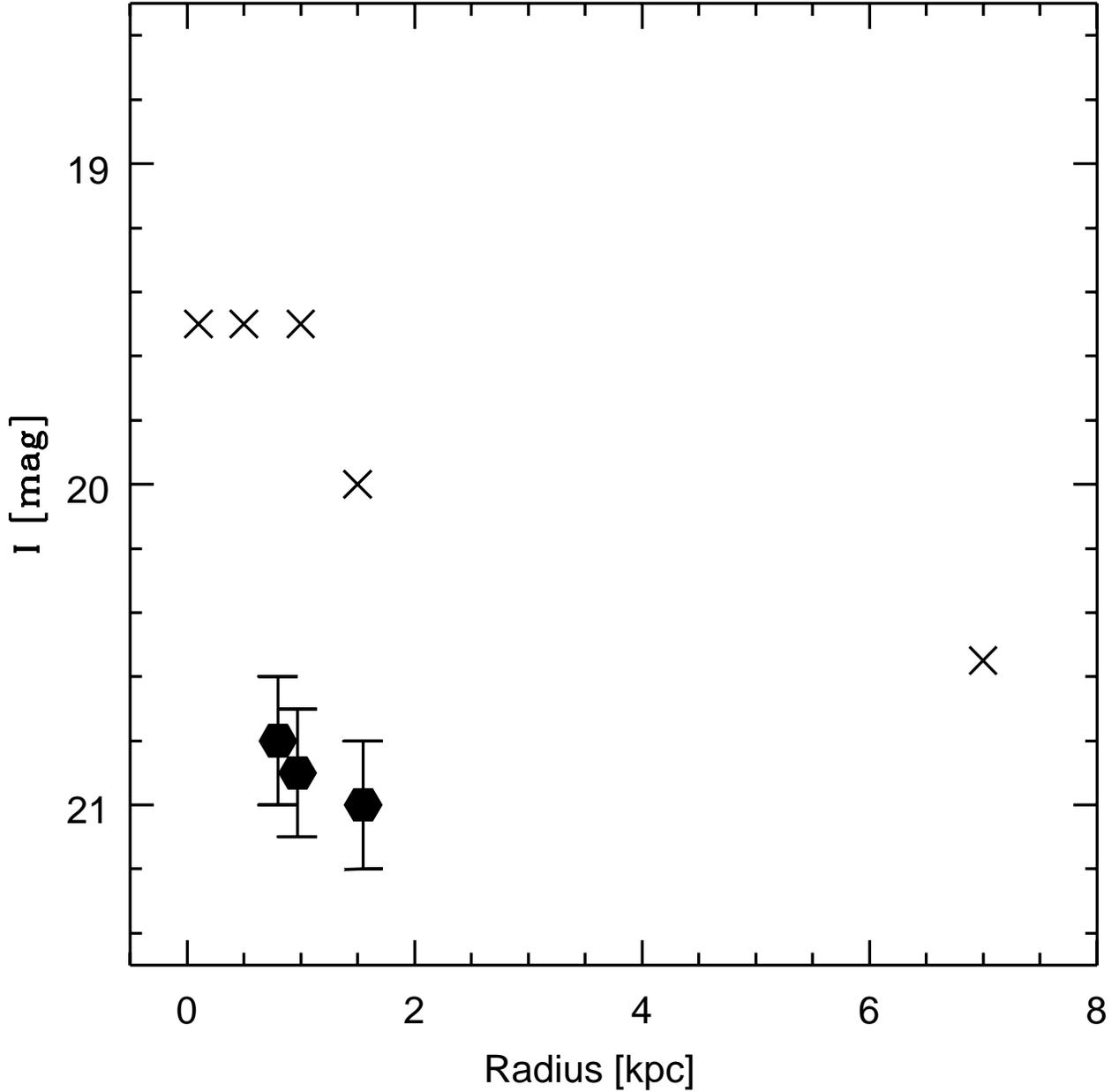,width=\the\hsize}}
\caption[]{\label{fig.diagram}   Comparison between   our  (PC)  data  and
previous analyses for the upper {\it envelopes} of the color-magnitude
diagrams.  Crosses  are for the HST  observations  of Rich  \& Mighell
(1995), ground-based   data of  Mould (1986)   and Mould  \&  Kristian
(1986). The hexagons show the measurements done in the present work. }
\end{figure}

\begin{figure}[h]
\centerline{
\psfig{bbllx=15mm,bblly=57mm,bburx=198mm,bbury=242mm,%
figure=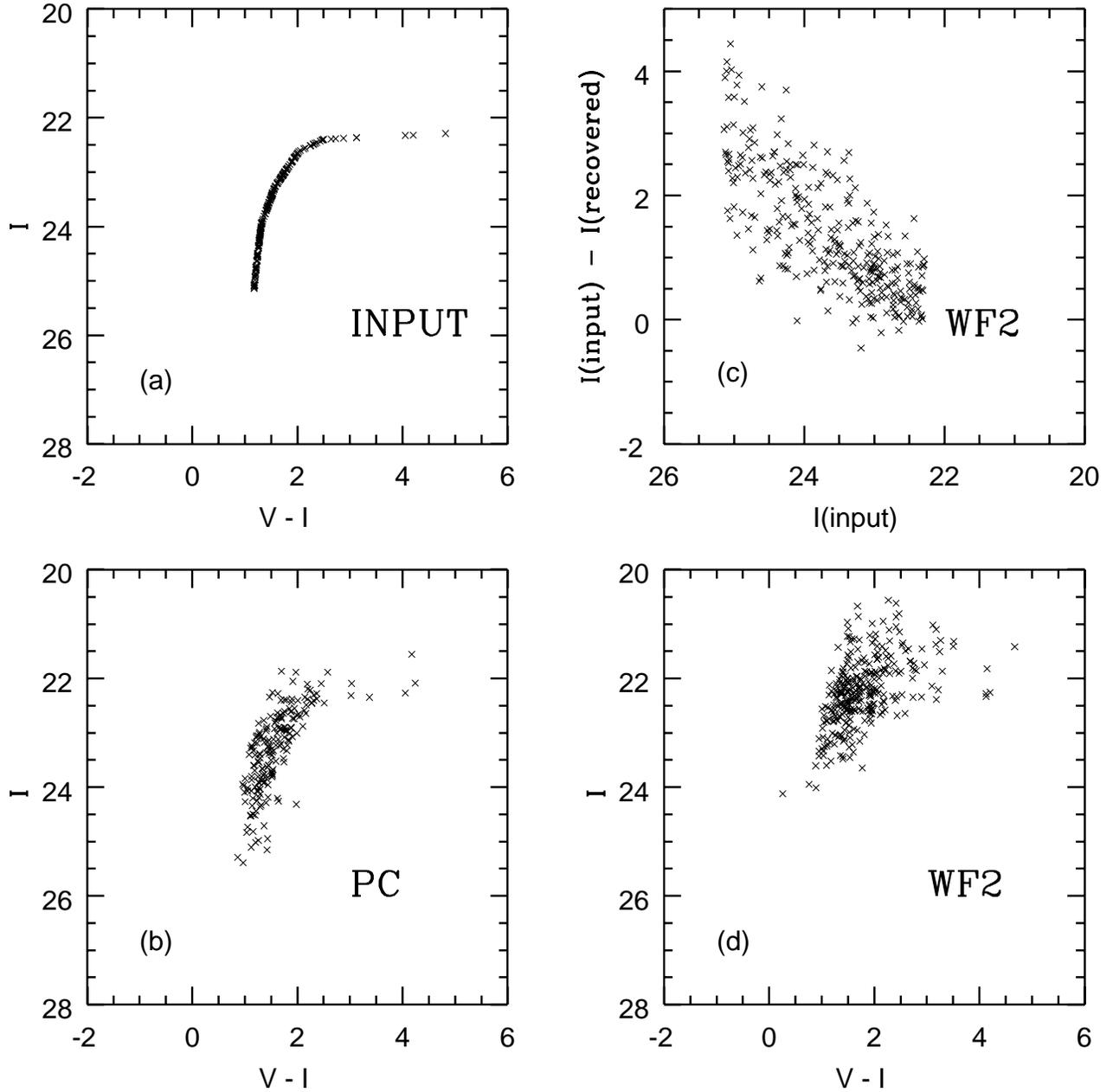,width=\the\hsize}}
\caption[]{\label{fig.diagram}   The  results  of the  artificial star
experiments.   The input   photometric   sequence is   shown in  Panel
(a). Panel (b) displays the recovered  photometry when the PC frame is
analysed. The error on the recovered  photometry when the WF2 frame is
considered  is illustrated in   Panel (c) and the resulting  recovered
color-magnitude diagram for the WF2 frame is shown in Panel (d). }

\end{figure}

\end{document}